# Revealing the Buried Metal-Organic Interface: Restructuring of the First Layer by van der Waals Forces

Margareta Wagner[1,2], Stephen Berkebile[1,3], Falko P. Netzer[1], Michael G. Ramsey[1]

[1] *Surface and Interface Physics, Institute of Physics, Karl-Franzens University, 8010 Graz, Austria*
[2] *Present address: Institut für Angewandte Physik, Technische Universität Wien, Wiedner Hauptstrasse 8-10/134, 1040 Vienna (Austria)*
[3] *Present address: U.S. Army Research Laboratory, Aberdeen Proving Ground, MD 21005, USA*

Corresponding author: *wagner@iap.tuwien.ac.at*

**Abstract**

Using molecular manipulation in a cryogenic scanning tunneling microscope, the structure and rearrangement of sexiphenyl molecules at the buried interface of the organic film with the Cu(110) substrate surface have been revealed. It is shown that a reconstruction of the first monolayer of flat lying molecules occurs due to the van der Waals pressure from subsequent layers. In this rearrangement, additional sexiphenyl molecules are forced into the established complete monolayer and adopt an edge-on configuration. Incorporation of second layer molecules into the first layer is also demonstrated by purposely pushing sexiphenyl molecules with the STM tip. The results indicate that even chemisorbed organic layers at interfaces can be significantly influenced by external stress from van der Waals forces of subsequent layers.

The interface between organic films and inorganic substrates is a prime determinant to the function of organic devices. As a consequence there has been a strong focus in the literature on the electronic structure and the electronic level alignment at the interface of π conjugated systems.[1-6] Yet of equal importance to device performance is the morphology and molecular and crystallite orientation. These geometric properties are also dependent on the interface on which the films grow.[7] Any control of bottom-up construction necessitates an understanding of the self-assembly process and, in turn, the interface on which it proceeds. Changes from those structures observed sequentially in layer-by-layer growth have been neglected because probing the buried interface is challenging, as it is essentially invisible to both bulk and surface sensitive techniques.

Unlike inorganic materials, the interfaces of organic materials are "soft". Both molecule/substrate and molecule/molecule interactions are dominated by relatively weak and directional van der Waals force, which results in a very delicate balance between intra- and inter-layer forces. Para-sexiphenyl (6P) on Cu(110) provides an excellent model system to investigate these phenomena as numerous experimental and theoretical studies of the geometric, and optical and electronic properties of both the first few 6P layers and thick films exist.[8-15] Moreover, 6P is tractable to scanning probe manipulation.[16] Recently, the growth and dynamics of 6P/Cu(110) have been investigated by photoemission electron microscopy (PEEM), providing an insight into growth mechanisms and dynamics of the first few layers.[17, 18] PEEM results indicate enhanced 1-dimensional diffusion of 6P molecules on formation of 3D crystallites that occurs after deposition of the third layer, which was postulated to be the result of a highly corrugated film forming on dewetting of metastable layers. Indeed STM images show a highly corrugated molecular film with troughs aligned in [1-10] direction.[18] Here the nature of the instigator for this corrugated film morphology has been explored using molecular manipulation in a cryogenic STM. We have been able to "dig" down with the STM tip through several molecular layers to the Cu interface and to image the first monolayer at the interface. It is shown that the otherwise dense and stable chemisorbed monolayer, consisting of flat lying molecules, is modified by the incorporation of edge-on molecules once further layers are deposited, effectively increasing the monolayer density. Moreover, these edge on molecules form semiperiodic chains parallel to the long molecular axis. We have also been able to simulate this incorporation by molecular manipulation.

**Results and Discussion**

In this work, a Cu(110) surface covered with 1.5 monolayers of 6P, deposited at 325 K, has been investigated in a low temperature (5K) STM.[19] Figure 1(a) gives an overview of the surface morphology. The first 6P wetting layer consists of flat and planar molecules forming a dense c(22×2) structure, where the molecules are aligned with their long molecular axis in the [1-10] direction.[10] For sub-monolayer exposures, the molecules repel each other due to Coulomb repulsion when the LUMO is occupied on hybridization. It is not clear whether the adsorption site changes on monolayer formation. Indeed, calculations for the c(22×2) structure could not energetically distinguish between molecules adsorbed on top of the atomic rows or in the troughs of the substrate.[14] The absorption site is determined with the aid of molecular manipulation (see below). Islands of the second layer grow from Cu step edges into the lower terrace areas (see middle part of Figure 1(a)). Figure 1(b) shows greater details of the second layer molecules that stack in the [001] direction and sit directly above those of the first layer. However, while the first layer molecules have co-planar phenyl rings and are in a flat adsorption geometry due to the strong interaction with the Cu substrate, the molecules in the second layer are less strongly bound and are twisted with phenyl rings tilted from the surface. This alternating tilt can be recognized in Figure 1(b) by the contrast modulation and the alternating brightness of individual rings along many of the molecules. The reduction in steric hindrance allows for a higher packing density and stronger intermolecular interactions. Yet although the second layer molecules have begun to approach their bulk configuration of twisted and tilted phenyl rings[20], the second layer spacing remains significantly larger than the bulk configuration due to interactions with the first layer. The second layer is always decorated by quasi-periodic bright chains of molecules running along the [1-10] direction. At room temperature these chains of molecules are very mobile, but at low temperature they become immobilized and can be imaged in the STM.

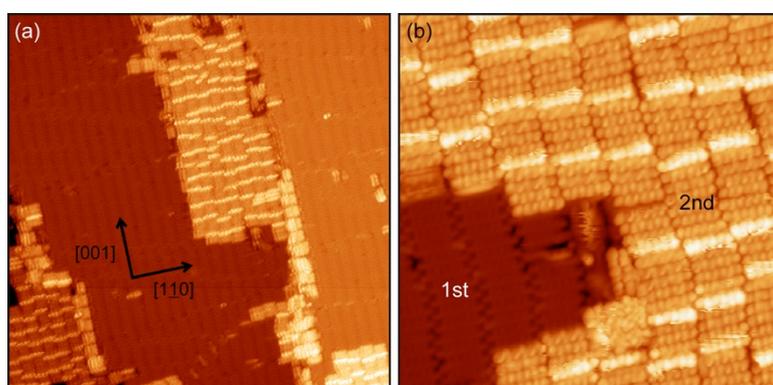

Figure 1: STM images of 1.5 monolayer of sexiphenyl deposited on Cu(110) at 325 K. (a) (800×800) Å$^2$; sample bias voltage $V_S = +1$ V; tunneling current $I_T = 0.1$ nA. (b) (200×200) Å$^2$; $V_S = +1$ V; $I_T = 0.1$ nA. First monolayer and second layer are indicated.

**Adsorption Site in the Sexiphenyl Monolayer on Cu(110)**

To obtain the adsorption site in the c(22×2) structure requires regions of bare Cu to be revealed by molecular manipulation. In order to clear the Cu surface of molecules adjacent to an area covered by a 6P monolayer, molecules from the upper terrace have been removed by pushing them down onto the lower terrace. The STM image of Figure 2(a, left panel) shows an area of the 6P covered Cu(110) surface prior to manipulation; the molecules that are subsequently removed are circled. The right panel of Figure 2(a) shows the same area after removal of 6P molecules, *i.e.*, a section of the bare Cu surface. Once the bare substrate area has been prepared, atomic resolution is required in the STM under conditions that do not disturb the adjacent molecules. Figures 2(b) and (c) displays STM images, where the dense Cu atomic rows in [1-10] direction as well as the monolayer molecules are well resolved. In the STM image of Figure 2(c), (bottom panel) a grid has been superimposed in such a way that the corners match the four-fold hollow sites of the Cu(110) surface. By extending the grid over the molecules one finds that the 6P molecules are adsorbed with their long axis center in between the Cu rows. Since one 6P molecule occupies the space of 11 Cu atoms in [1-10] direction, spacing of the phenyl rings and the underlying Cu atoms cannot be identical. Analysis of the registry in the [1-10] direction reveals that the 6P in the monolayer occupies the same adsorption site as an isolated 6P molecule:[10] the phenyl rings 1 and 4 are located above long bridge positions (marked by arrows in Figure 2(c), the positions of phenyl rings 2, 3, 5 and 6 are slightly off-center of four-fold hollow sites. The lower part of Figure 2(d) gives a model of the monolayer structure, with the van der Waals size of 6P indicated.

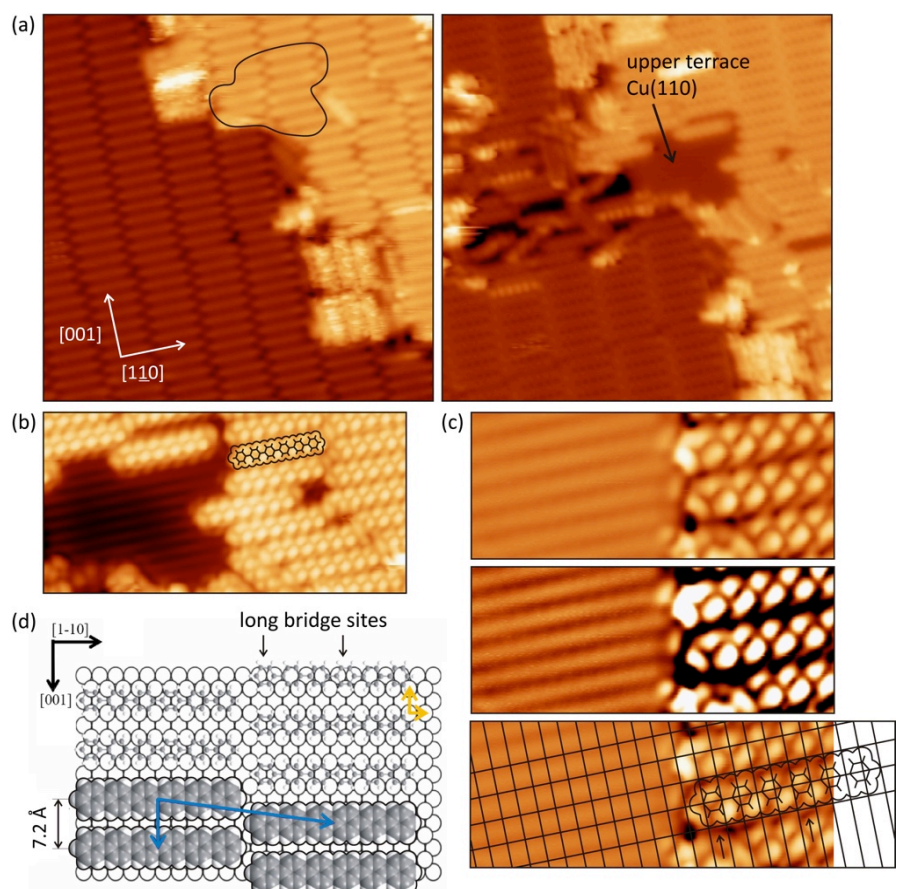

Figure 2: Digging down to the Cu(110) substrate. (a) STM images of the 6P covered Cu(110) surface before (left) and after (right) removal of 6P molecules by STM tip manipulations ((200×200) Å$^2$; $V_S$ = -1 V; $I_T$ = 50 pA). (b) STM image displaying atomic resolution of both Cu substrate and 6P molecules: 6P adsorbs in the troughs of the Cu(110) surface. (100×51) Å$^2$; $V_S$ = -65 mV; $I_T$ = 0.84 nA. (c) High resolution image of the Cu(110) surface displayed in different color contrasts and (bottom) superimposed with a grid of the Cu(110) lattice. (50×21) Å$^2$; $V_S$ = -11 mV; $I_T$ = 0.84 nA). (d) Model of the Cu(110) surface with 6P molecules in the monolayer configuration.

**Bilayer Structure and Monolayer Reconstruction**

In the following, the attention will be focused on the nature of the chains of bright molecules in [1-10] direction in the second layer and their relationship to the underlying monolayer.

The structure of the first monolayer underneath the second-layer islands is investigated by partially removing second-layer molecules step by step with the STM tip as shown in Figure 3. The removal of the two bright molecules (indicated with dots in panels (1) and (2)) reveals dark channels associated with vacancies in the second layer (marked by the arrows). From panels (2) to (4), the molecules of the second layer adjacent to these troughs (in [001] direction) have been removed. It is clear from panels (3) and (4) that below the vacancies of the second layer, *i.e.*, below the bright second layer molecules that have been removed, molecules of the first layer with somewhat brighter contrast are located. This contrast suggests a different configuration of these molecules as demonstrated below. A second pair of bright molecules of the second layer (see dots in panel (4)) is removed in panels (4) to (5). In the latter, four such molecules (below the troughs) with brighter contrast are apparent. In panel (6), the three second layer molecules between the troughs have also been removed. Note the somewhat different STM contrast in panels (1-3) and (4-6) due to the change of bias polarity between the two sets. This change causes the appearance of two linear features for one edge-on molecule due to the imaging of different electronic orbitals in the second set of panels. Linescans (not shown here) show the height of the decorating molecule to be 3.0 Å, the second layer 1.5 Å, and the species revealed after removing the decorating molecules to be 1 Å relative to the top of the monolayer. The height of the revealed molecule suggests that these molecules underneath the decorating molecules are edge-on.

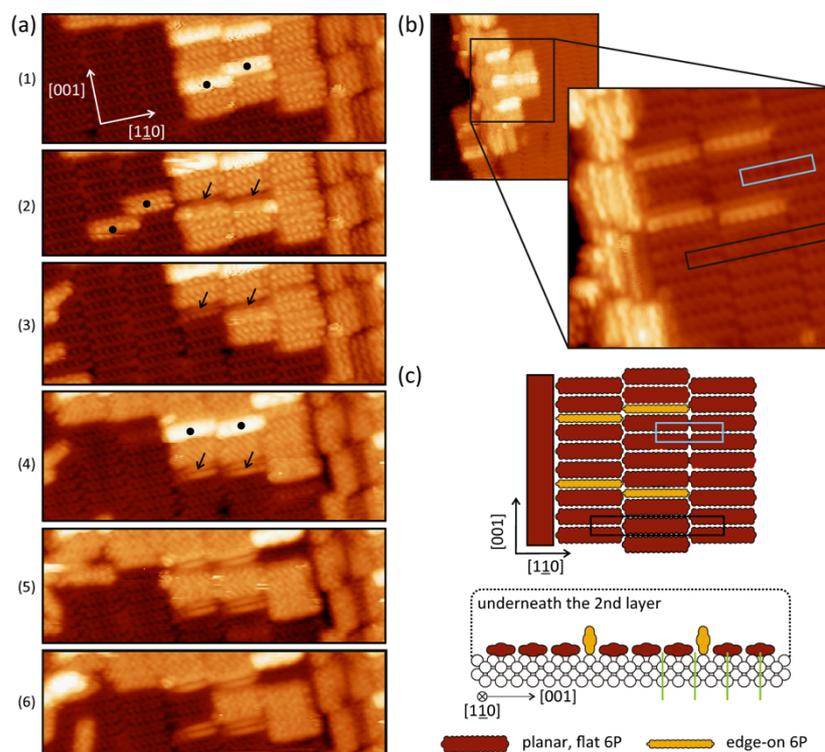

Figure 3: Digging down to the first layer with the STM tip by removing molecules of the second layer step by step. (a) STM images in panels 1–3: (200×35/28/33) Å$^2$; $V_S$ = -1 V; $I_T$ = 67 pA); panels 4–6: (200×35/31/28) Å$^2$; $V_S$ = +1 V; $I_T$ = 67 pA). The position of troughs in the second layer are marked by arrows. (b) STM images of a small second layer island before (upper left) and after its removal (lower right). The unit cell of the monolayer (black) and the local unit cell at the domain boundary

(blue) are indicated. (upper left panel: (200×200) Å$^2$; $V_S$ = -1 V; $I_T$ = 47 pA; lower right panel: (100×100) Å$^2$; $V_S$ = -1.6 V; $I_T$ = 28 pA). (c) Schematic top and side views of the rearranged monolayer corresponding to the STM image in (b, lower right panel).

A close inspection of the first monolayer 6P reveals an anti-phase domain boundary in [001] direction between the two brighter molecules (formerly below the troughs) and the surrounding first layer molecules, as illustrated in Figures 3(b) and (c). In Figure 3(b), a small second layer island (upper left panel) has been removed (lower right panel), leaving two pairs of brighter first layer molecules. The c(22×2) unit cell of the first layer (black) and the local unit cell of the domain boundary (blue) are indicated in (b), whereas the situation is schematically depicted in Figure 3(c). The antiphase domain boundary is the result of molecules shifted by one Cu lattice parameter along [001]; note that the monolayer periodicity is two Cu rows in the [001] direction, which allows for a shift of one Cu row by an edge-on molecule. It is important to realize at this point that the area underneath the second layer island exhibits a higher density of molecules, which is not possible if all first layer molecules remained in their flat lying configuration. Hence it is suggestive to assume that, if one additional molecule is incorporated into the monolayer, two molecules in an edge-on configuration must be created: each of the latter occupy half the area of a flat lying molecule. The increase in density of the reconstructed monolayer is 5:4 compared to the perfect c(22×2) monolayer. The shift of one 6P molecule lattice site also causes a shift in domain and an antiphase boundary.

This conjecture can be substantiated experimentally by molecular manipulation with the STM, whereby molecules can be purposely pushed into the first layer. The process is demonstrated in Figure 4(a), where a molecule of the second layer is first separated from an island in panels (1)-(3), and this is then followed by the incorporation into the first layer by pushing it in [1-10] direction, panel (4). Finally, in panel (5), the originally twisted molecule of the second layer becomes a planar and flat molecule of the first layer, whereas the neighboring molecules, originally planar and flat, are now brighter in contrast to surrounding first layer molecules. Fig 4(b) explains the spatial arrangement of the molecules during incorporation with a schematic representation. When the additional 6P molecule (white) is added to the first layer, three molecules occupy the area where formerly only two molecules were. The incorporation appears to be a local effect that leaves the molecules beyond nearest neighbors undisturbed, but the nearest neighboring molecules rotate and flip edge-on (yellow). Once a molecule has been incorporated it is very easy to move the position of the edge on molecules in the [001] direction. This procedure is illustrated in Figure 4(c) by pushing such an edge-on molecule perpendicular to its long axis (as indicated by the arrow). A domino effect results that flips the edge-on molecule back into the flat configuration at the expense of forcing the next molecule edge-on (lower panel). The lower panel of Figure 4(c) reveals also an anti-phase domain between the edge-on molecules (blue rectangle), which is consistent with the findings in the first layer underneath the second layer islands (Figure 3(b)).

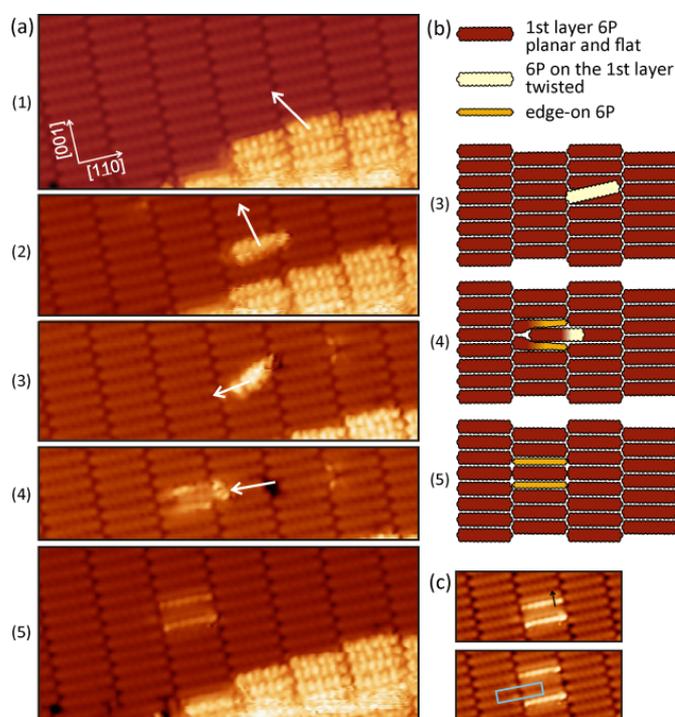

Figure 4: Incorporation of a 6P molecule into the monolayer by molecular manipulation – pushing it with the STM tip in [1$\underline{1}$0] direction. (a) STM images with the direction of the manipulation indicated by arrows. (100×46/32/32/25/46) Å$^2$; $V_S$ = -1 V; $I_T$ = 48 pA. (b) Cartoons of the crucial steps during the incorporation process. Panel numbers correspond to those in (a). (c) STM images demonstrating the domino action: flipping molecules from flat to edge-on and vice-versa with the STM tip. (100×43) Å$^2$; $V_S$ = +1 V; $I_T$ = 48 pA.

The scenario of the rearrangement of the buried monolayer at the metal interface is illustrated by the schematics of Figure 5. After formation of the c(22×2) monolayer of flat-lying 6P molecules (step 1), the second layer becomes populated with flat-lying but twisted molecules (step 2). At a critical second layer density, additional molecules are pushed into the first layer, causing adjacent molecules to flip edge-on to provide the necessary extra space (steps 3 and 4). Eventually, the edge-on molecules become decorated by second layer molecules, which show up with the brighter contrast in STM images.

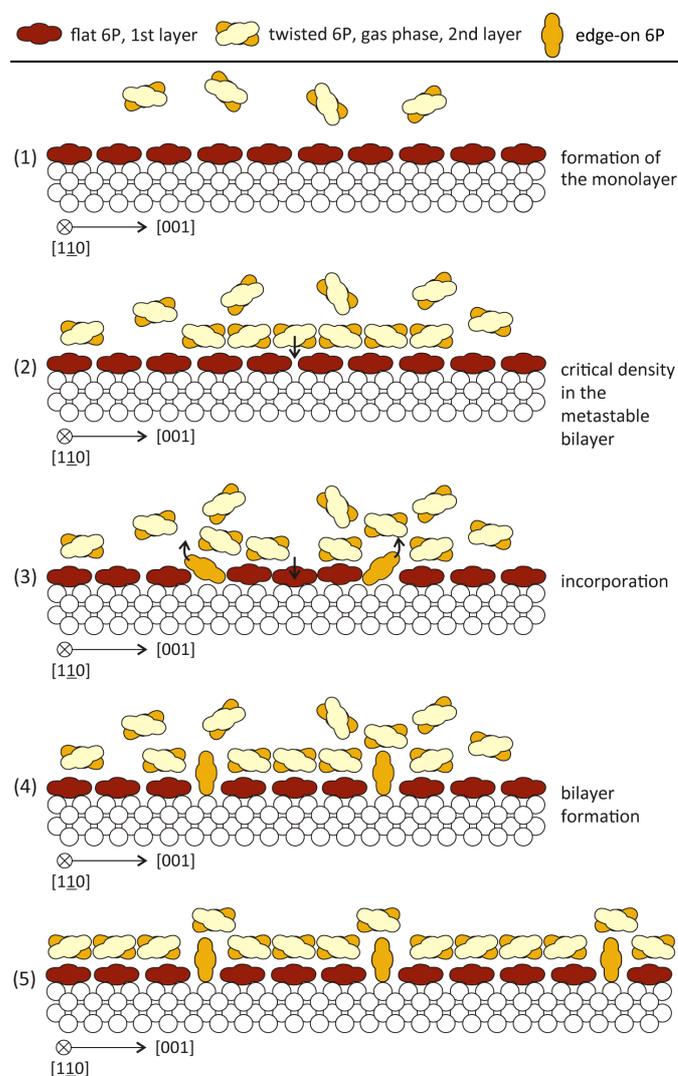

Figure 5: Schematic illustration of the growth scenario of 6P molecules on Cu (110) and of the rearrangement of the buried monolayer at the metal interface with steps 1 through 5. Molecules of the monolayer and edge-on molecules are planar; molecules of the gas phase and the second layer have twisted phenyl rings.

**Channel Superstructure**

When the monolayer is covered by the second layer (together, the bilayer), the edge-on molecules are decorated by bright molecules in linear chains running in [1$\bar{1}$0] direction. The STM image of the bilayer also shows that the bright linear chains of 6P molecules run across the whole island with no open ends, *i.e.*, usually chains do not stop in the middle of the island (see left panel in Figure 6(a)). In the following, the spacing between two chains is named "channel".

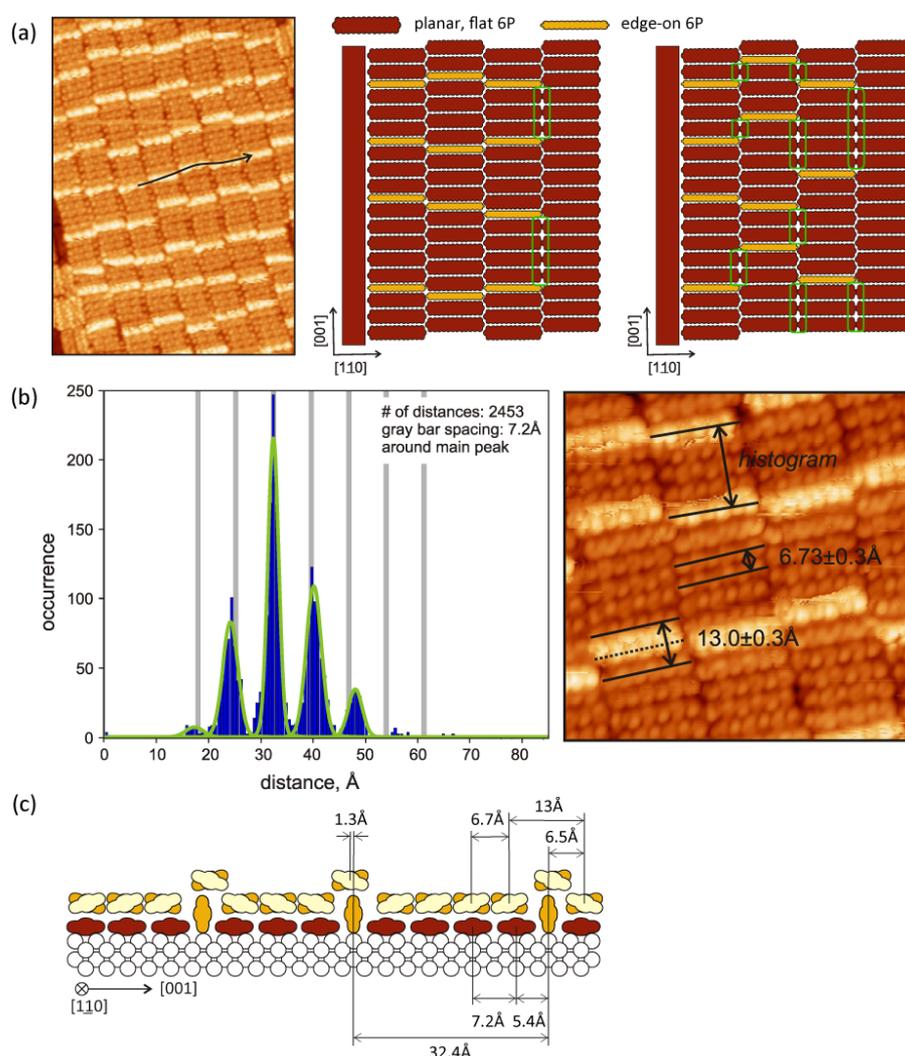

Figure 6: (a) Formation of linear chains to avoid anti-phase domain boundaries in the monolayer. STM image of the bilayer ((210×282) Å$^2$; V$_S$ = -1 V; I$_T$ = 0.1 nA). Cartoons: Schematic drawings of two extreme scenarios for the arrangement of the edge-on molecules with the antiphase domains indicated. (b) (left) Statistics of the spacing of the bright second layer molecules in [001] direction. (right) The second layer with distances inserted. ((100×100) Å$^2$; V$_S$ = -1.1 V; I$_T$ = 0.1 nA). (c) Schematic illustration of the 6P bilayer on Cu (110).

We propose that the formation of these chains of edge-on molecules (and consequently the bright molecules in the second layer which sit on top of them) minimizes the antiphase domain boundaries created when a molecule is incorporated. Figure 6(a) presents two scenarios for the arrangement of a reconstructed monolayer, where the edge-on 6P are in chains (middle panel) or are randomly distributed (right panel). In the random case antiphase domain boundaries are omnipresent, which is energetically unfavorable due to steric hindrance and intermolecular repulsion of the hydrogen atoms. Minimizing these boundaries is achieved if the edge-on molecules arrange in next neighboring sites (with respect to the [1$\underline{1}$0] direction and one Cu lattice constant shifted up or down in [001] direction). In this arrangement, an antiphase domain boundary will occur only at the ends of every second channel (see middle panel of Figure 6(a)), and when a terrace is completely covered there will be no antiphase domains.

To better understand the stacking of the molecules in the second layer, the spacing of the molecules within a channel and the spacing of the channels has been analyzed in detail. The spacing of molecules within a channel has been measured to be (6.73 ± 0.3) Å on average. The distance between the molecules of the second layer on either side of the molecule sitting on-top of the edge-on 6P is (13.0 ± 0.3) Å (indicated in the STM image of Figure 6(b)). This implies a value of only (6.5 ± 0.3) Å for the distance between the center of the edge-on molecule (dotted line in the STM image of Figure 6(b)) and the next neighboring molecule in the second layer (solid lines). The number of molecules in bilayer channels is always the same as the number in the underlying monolayer although the bilayer is strictly incommensurate (6.7 Å *versus* 7.2 Å lateral spacing). The lateral spacing of molecules in the bulk sexiphenyl crystal is 5.6 Å and a similar structure with a 5.55 Å spacing is observed for 6P monolayer islands on the oxygen passivated Cu(110)-(2×1)O surface [10, 11]. Significantly, the second layer spacing of 6.7 Å is much larger than the bulk spacing, indicating the influence of the 7.2 Å monolayer.

The spacing between the edge-on 6P in the [001] direction is semi-periodic and in fact a semi-commensurate superstructure in this direction, as the following analysis reveals. The distance between second layer molecules decorating the edge on molecules has been statistically evaluated by computationally fitting their positions from height profiles taken along [001] direction. The histogram in Figure 6(b) depicts the distances between these bright molecules, as indicated in the STM image on the right hand side. The distribution consists of a sharp main maximum at 32.39 Å and side maxima. This distance is roughly determined by the number of molecules between the edge-on molecules in [001] direction since the number of molecules in the monolayer and second layer is the same, as indicated by the histogram peak spacing close to 7.2 Å. This most preferred situation corresponds to a spacing of 9 Cu rows or 4.5 monolayer 6P, which is 32.4 Å, which means that the underlying edge-on molecules are most frequently separated by four flat-lying molecules.

The side maxima to the main 32.39 Å maximum are located at 17.3 Å, 24.18 Å, 40.10 Å, and 48.05 Å, with an average separation between the maxima of 7.96 Å. Further, overall the maxima are slightly skewed in occurrence towards a larger spacing from the main maximum. A Gaussian fit of all separation values together leads to a center of 33.44 Å, which is close to the spacing of six second layer molecules within a channel of the second layer (33.65 Å). The results of the statistical analysis and molecular spacing are summarized in the schematic of Figure 6(c).

Interestingly, these distances indicate that the bright 6P of the second layer are not perfectly centered on the edge-on molecules, but either shifted up or down in [001] direction. The off-centered position appear to be related to the number of molecules in the channel above and below, respectively, in such a way that they are always shifted by 1.3 ± 0.5 Å away from the channel with more molecules. The direction of the shift is randomly up or down by this value when equally wide channels are on either side. The histogram displays this behavior in the position of the side maxima. They are found to be larger than a spacing of 7.2 Å from the main maximum for those above the main maximum and smaller for those below, rather than spread broadly or displaying a double maximum around a peak spacing of 7.2 Å.

The geometry between the two layers strives to maximize π-system overlap, and consequently the van der Waals forces between the layers. However, while the first layer is constrained by the substrate to a lateral spacing of 7.2 Å, the molecules in the second layer will be subject to a mutual attraction that tends to pull them towards the closer packing of the 6P crystal (5.6 Å). These competing forces lead to a strain in the second layer. The difference in spacing between the monolayer and the molecules within the channel leads to a maximum strain in the [001] direction dependent on the number of molecules within the channel. Assuming that the molecules are centered within the channel, the difference in lateral molecular positions between the two layers at the outermost molecules in the channel is $d = (a - b)(\frac{m-1}{2})$, where $a$ = 7.2 Å, $b$ = 6.73 Å, and $m$ is the number of molecules (in either the second or first layer). The maximum percent mismatch from $a$ = 7.2 Å for the observed spacing maxima corresponding to three, four, five and six molecules is 6.5%, 9.8%, 13.1% and 16.3%, respectively. It is this energy balance between the layers and within the second layer that results in the distribution of edge-on molecule spacings distributed around a maximum of four molecules. Above 16.3%, the strain induced between the layers far exceeds the energy gain from the van der Waals forces leading to very few instances of higher spacings.

Edge-on 6P molecules exist only underneath the second layer, but have not been observed in the uncovered monolayer on Cu(110), which demonstrate the importance of the second layer in modifying the balance of interactions between the molecules of the first layer and the substrate. In contrast to the strongly bound 6P monolayer on Cu(110), the more weakly bound system of 6P on Au(111) and HOPG are found to transition within the monolayer with increasing coverage (and intermolecular interaction). On Au(111), a transition from a uniform adsorption geometry to a tilted, flat alternating structure that more closely replicates the herring-bone structure of the bulk crystal occurs[21, 22] and, on HOPG, initially flat molecules reorganize as a layer to a standing, bulk-like configuration.[23] Furthermore, upon thermal desorption of the second layer, the c(22×2) structure of the monolayer is regained. This indicates that the incorporation of additional molecules into the first layer requires the chemical pressure of the second layer *via* van der Waals interactions. The average separation of edge-on molecules by four flat molecules indicates that the molecule coverage at the interface can be increased by ~17%. The additional adsorption energy of edge-on molecules thus compensates for the formation of antiphase domain boundaries and the overall less-favorable adsorption geometry that discourage spacings below three molecules.

**Conclusion**

We have shown that the van der Waals pressure of additional layers may induce a reconstruction of the first layer at the interface. Sexiphenyl on Cu(110) is a model system for demonstrating that organic layer interfaces are not rigid but soft and flexible. Using molecular manipulation with the STM tip, we have revealed this structural rearrangement at the buried organic-metal interface with layers of 6P on Cu(110). In the present case, the first monolayer of 6P is elastic in the [001] direction, and the pressure from a second layer of molecules is sufficient to press additional molecules towards the metal interface, inducing a domino-like reaction and a rearrangement with edge-on molecules in the interfacial layer. The repulsive interaction of 6P molecules in the first layer, which prevents island formation in the monolayer, competes with the attractive interaction of the tilted/twisted molecules in the second layer to reconstruct the first layer and form a complex semi-commensurate superstructure dependent on the balance of forces within and between each layer. This superstructure is both a step towards adopting the bulk structure with continued organic film growth and determines the subsequent growth kinetics by forming one dimensional channels for diffusion.

The phenomena can be expected to be found in other aromatic systems that pack in a herring bone structure if their first layer adsorbs with their aromatic planes parallel to the surface. If the bond to the substrate is too weak, as is the case for sexiphenyl on the oxygen passivated surface, the phenomena does not occur as the first layer can adopt a geometry near to that of the organic crystal. At present, theory has not been demonstrated to reliably cope with such a large system (>1000 atoms) dominated by van der Waals interactions to the accuracy and length scale required to capture these delicate interlayer interactions. Although density functional theory can incorporate van der Waals interactions,[24] accurate methods have usually been prohibitively complex.[25] A DFT method that incorporates van der Waals forces has been used to describe the modification of surface bonding distance and charge redistribution that occurs in a first molecular layer when a second layer of a different molecule is present,[26] yet it remains to be seen whether new advances in first principles calculations will be able to describe structures that develop with longer range order that includes interactions tens of molecules. However, evidence for the generality may be sought in complex thermal desorption spectra that appear for small aromatics such as benzene[27] or bithiophene[28] on certain metal surfaces that have been attributed to metastable layers, and in the complex tetracene structures formed on Ag(111).[29]

**Experimental**

The experiments have been carried out in a three-chamber UHV system with a base pressure below $5\times10^{-11}$ mbar, equipped with a low temperature STM (CreaTec, Germany) operating at 5 K, a home-made molecule evaporator, a quartz micro balance to monitor the molecule deposition rate, LEED, and the usual facilities for sample manipulation and cleaning.[19] Electrochemically etched W wire of 0.3 mm diameter has been used for STM tips, which have been cleaned *via* electron bombardment in UHV and by in-situ treatments in the STM stage such as voltage pulses. STM images have been recorded in constant current mode, with bias voltages applied to the sample. Manipulations of molecules with the STM tip have been performed by pulling or pushing molecules several 10 Å across the surface with a lateral speed of ~190 Å/s. Typical STM manipulation parameters for moving 6P on the Cu(110) surface, on the 6P monolayer, and on the second layer, as well as typical imaging conditions are summarized in Table 1.

|  | On Cu(110) along [1-10] | On the 1st and 2nd layer 6P | Imaging |
|---|---|---|---|
| V (V) | 0.03 | 0.9-0.1 | +2 to -2 |
| I (nA) | 100 | 50-100 | 0.01-1 |
| R (MΩ) | 0.3 | 7-1 | ~1000 |

Table 1: STM set point (bias voltage V and tunneling current I) for manipulating individual sexiphenyl molecules on Cu(110), and the first and second layer 6P, and typical imaging conditions used on this surface.

The Cu(110) surface has been cleaned by cycles of $Ar^+$ bombardment (0.8-1 keV) and subsequent heating to 823 K. Surface order and cleanliness has been checked by LEED and STM. In particular, small amounts adsorbed oxygen will induce a different alignment of 6P molecules.[10,30].

**Acknowledgments**

This work has been supported by the by the ERC Advanced Grant "SEPON".